\documentclass[aps,showpacs,preprintnumbers,amsmath, amssymb]{revtex4}

\usepackage{float}
\usepackage{graphics,epsfig}
\usepackage{graphicx}
\usepackage{dcolumn}
\usepackage{bm}

\begin{document}
\baselineskip=0.8 cm

\title{{\bf Scalarization of compact stars in the scalar-Gauss-Bonnet gravity}}
\author{Yan Peng$^{1}$\footnote{yanpengphy@163.com}}
\affiliation{\\$^{1}$ School of Mathematical Sciences, Qufu Normal University, Qufu, Shandong 273165, China}

\vspace*{0.2cm}
\begin{abstract}
\baselineskip=0.6 cm
\begin{center}
{\bf Abstract}
\end{center}

We study scalarization of horizonless neutral
compact reflecting stars. In our model, the scalar hair can be
induced by the coupling of static scalar fields to the Gauss-Bonnet invariant.
We analytically obtain lower bounds on the
coupling parameter. Below the bound, the static scalar hair cannot form.
And above the bound, we numerically get the discrete
coupling parameter that can support scalar hairs outside stars.
We also disclose effects of model parameters on the discrete
coupling parameter.

\end{abstract}

\maketitle
\newpage
\vspace*{0.2cm}

\section{Introduction}

The recent direct detection of gravitational waves provided
the strong evidence that extremely compact black holes
exist in nature \cite{BP1,BP2,BP3}.
Astrophysical black holes were usually believed to
be endowed with an absorbing horizon. Interestingly, however,
some candidate quantum-gravity models suggested that
quantum effects may prevent the formation of classical horizons
\cite{UH1,UH2,UH3,UH4,UH5}.
An intriguing idea is replacing the horizon
by a reflecting wall outside
(but extremely close to) the gravitational radius
\cite{RS1,RS2,RS3,RS4,RS5,RS6,RS7}.
From astrophysical and theoretical aspects,
it is very interesting to search for similarities
and differences between such horizonless reflecting compact
stars and classical absorbing black holes.

For classical black holes, one famous property is the no hair theorem,
see references \cite{Bekenstein}-\cite{sn4} and reviews \cite{Bekenstein-1,CAR}.
The original no hair theorem states that static scalar fields cannot
exist outside asymptotically flat black holes.
Nevertheless, the scalar hair can form when violating some of their
assumptions, such as considering stationary scalar fields or
enclosing the black hole in a box \cite{vn1}-\cite{vn10}.
Another well studied model is considering static scalar fields
non-minimally coupled to the Gauss-Bonnet invariant,
which allows the existence of thermodynamically stable
hairy black holes \cite{SGB1,SGB2,SGB3,SGB4,SGB5,SGB6,SGB7}.
This scalar-Gauss-Bonnet gravity attracted lots of attentions
and more complete models were constructed
\cite{SGB8,SGB9,SGB10,SGB11,SGB12,SGB13,SGB14,SGB15,SGB16}.

For horizonless compact reflecting stars \cite{RS1,RS2,RS3,RS4,RS5,RS6,RS7},
intriguingly, such no scalar hair behavior also appears.
Hod firstly showed that massive static scalar fields cannot
exist outside asymptotically flat neutral horizonless compact stars
with a scalar reflecting surface \cite{Hod-1}.
For massless static scalar fields outside such
asymptotically flat neutral horizonless reflecting compact stars,
the no hair theorem was firstly analysed in \cite{Hod-5}
and another proof was provided based on Cauchy-Kowalevski theorem \cite{Yan Peng-1}.
In the asymptotically dS gravity, static scalar hairs still cannot
form outside the neutral horizonless reflecting compact stars \cite{SBS}.
With field-curvature couplings, there is also no static hair theorem
for the neutral horizonless reflecting stars \cite{Hod-5,Hod-6}.
It seems that no static hair behavior is a very general property
in the background of neutral horizonless compact reflecting stars,
for extended discussion in charged backgrounds see \cite{Hod-2}-\cite{BKM}.
On the other side, as mentioned in the front paragraph, black holes can be hairy
in the scalar-Gauss-Bonnet gravity.
Along this line, it is interesting to examine whether the
scalar-Gauss-Bonnet coupling can induce scalar hairs
outside neutral horizonless compact reflecting stars.

This work is organized as follows.
We firstly construct a system with a scalar field coupled to the Gauss-Bonnet invariant
in the background of a neutral horizonless compact reflecting star.
Then we analytically obtain a lower bound on the coupling parameter,
below which there is no hair theorem.
And above the bound, we numerically get discrete coupling parameters
that can support horizonless neutral hairy compact stars.
At last, we give the main conclusion.

\section{Lower bounds on the scalar-Gauss-Bonnet coupling parameter}

In this paper, we consider the coupling of massive scalar fields
to the Gauss-Bonnet invariant in the asymptotically flat gravity.
The general Lagrange density reads \cite{SGB1,SGB2,SGB3,SGB4,SGB5,SGB6,SGB7}
\begin{eqnarray}\label{lagrange-1}
\mathcal{L}=R-|\nabla_{\mu} \psi|^{2}-m^{2}\psi^{2}+f(\psi)\mathcal{R}_{GB}^{2},
\end{eqnarray}
where R is the Ricci scalar curvature, $\psi(r)$ is the scalar field with mass m,
$\mathcal{R}_{GB}^{2}$ is the Gauss-Bonnet invariant in the form
$\mathcal{R}_{GB}^{2}=R_{\mu\nu\rho\sigma}R^{\mu\nu\rho\sigma}-4R_{\mu\nu}R^{\mu\nu}+R^2$
and $f(\psi)$ represents the coupling function.
Neglecting backreaction of scalar fields,
there is $\mathcal{R}_{GB}^{2}=\frac{48M^2}{r^6}$.
In the linear limit, without loss of generality,
one can put the coupling function in a simple quadratic
form $f(\psi)=\eta\psi^2$ with $\eta$ as the coupling
parameter \cite{SGB3,SGB4}.

In the Schwarzschild coordinates, the spherically symmetric
spacetime is described by \cite{SGB4}
\begin{eqnarray}\label{AdSBH}
ds^{2}&=&-g(r)dt^{2}+\frac{dr^{2}}{g(r)}+r^{2}(d\theta^{2}+sin^{2}\theta d\phi^{2}).
\end{eqnarray}
The metric outside the star is a Schwarzschild type solution
$g(r)=1-\frac{2M}{r}$, where M is the star mass.
We define the radial coordinate $r=r_{s}$ as the star surface radius.
The horizonless condition of the compact star requires the relation $r_{s}>2M$.

We take the static scalar field only depending on the radial
coordinate in the simple form $\psi=\psi(r)$.
After varying the action, the scalar field equation of motion is
\begin{eqnarray}\label{BHg}
\psi''+(\frac{2}{r}+\frac{g'}{g})\psi'+(\frac{\eta\mathcal{R}_{GB}^{2}}{g}-\frac{m^2}{g})\psi=0
\end{eqnarray}
with $g=1-\frac{2M}{r}$ and $\mathcal{R}_{GB}^{2}=\frac{48M^2}{r^6}$.

We assume that the scalar field vanishes at the star surface.
The bound-state (spatially localized) massive scalar fields are
characterized by asymptotically decaying behaviors
$\psi(r\rightarrow \infty)\sim \frac{1}{r}e^{-mr}$.
So the scalar field boundary conditions are
\begin{eqnarray}\label{InfBH}
&&\psi(r_{s})=0,~~~~~~~~~\psi(\infty)=0.
\end{eqnarray}

With a new function $\tilde{\psi}=\sqrt{r}\psi$,
one can rewrite the differential equation (3) into
\begin{eqnarray}\label{BHg}
r^2\tilde{\psi}''+(r+\frac{r^2g'}{g})\tilde{\psi}'+(-\frac{1}{4}-\frac{rg'}{2g}+\frac{\eta\mathcal{R}_{GB}^{2}r^2}{g}-\frac{m^2r^2}{g})\tilde{\psi}=0.
\end{eqnarray}

According to boundary conditions (4), the new function $\tilde{\psi}$ satisfies
\begin{eqnarray}\label{InfBH}
&&\tilde{\psi}(r_{s})=0,~~~~~~~~~\tilde{\psi}(\infty)=0.
\end{eqnarray}

The function $\tilde{\psi}$ must possess one extremum point $r=r_{peak}$
between the star surface $r_{s}$ and the infinity boundary.
If $r_{peak}$ is a positive maximum extremum point, there are relations
\begin{eqnarray}\label{InfBH}
\tilde{\psi}(r_{peak})>0, \tilde{\psi}'(r_{peak})=0, \tilde{\psi}''(r_{peak})\leqslant0,
\end{eqnarray}
otherwise it will be negative minimum extremum point satisfying
\begin{eqnarray}\label{InfBH}
\tilde{\psi}(r_{peak})<0, \tilde{\psi}'(r_{peak})=0, \tilde{\psi}''(r_{peak})\geqslant0.
\end{eqnarray}

In summary, at the extremum point $r=r_{peak}$, the scalar field is characterized by
\begin{eqnarray}\label{InfBH}
\{ \tilde{\psi}\neq 0,~~ \tilde{\psi}'=0~~and~~\tilde{\psi} \tilde{\psi}''\leqslant0\}~~for~~r=r_{peak}.
\end{eqnarray}

Multiplying both sides of (5) with $\tilde{\psi}$,
we arrive at the equation
\begin{eqnarray}\label{BHg}
r^2\tilde{\psi}\tilde{\psi}''+(r+\frac{r^2g'}{g})\tilde{\psi}\tilde{\psi}'+(-\frac{1}{4}-\frac{rg'}{2g}
+\frac{\eta\mathcal{R}_{GB}^{2}r^2}{g}-\frac{m^2r^2}{g})\tilde{\psi}^2=0.
\end{eqnarray}

Relations (9) and (10) yield the inequality
\begin{eqnarray}\label{BHg}
-\frac{1}{4}-\frac{rg'}{2g}+\frac{\eta\mathcal{R}_{GB}^{2}r^2}{g}-\frac{m^2r^2}{g}\geqslant0~~~for~~~r=r_{peak}.
\end{eqnarray}

It can be transformed into
\begin{eqnarray}\label{BHg}
(m^2-\eta\mathcal{R}_{GB}^{2})r^2g\leqslant -\frac{rgg'}{2}-\frac{1}{4}g^2~~~for~~~r=r_{peak}.
\end{eqnarray}

Since $r\geqslant r_{s}> 2M$, we have
\begin{eqnarray}\label{BHg}
g=1-\frac{2M}{r}=\frac{1}{r}(r-2M)>0,
\end{eqnarray}
\begin{eqnarray}\label{BHg}
rg'=r(1-\frac{2M}{r})'=\frac{2M}{r}> 0.
\end{eqnarray}

From (12-14), one deduces that
\begin{eqnarray}\label{BHg}
m^2-\eta\mathcal{R}_{GB}^{2}<0~~~for~~~r=r_{peak}.
\end{eqnarray}

According to (15), there are relations
\begin{eqnarray}\label{BHg}
\eta> \frac{m^2}{\mathcal{R}_{GB}^{2}}=\frac{m^2r_{peak}^6}{48M^2}\geqslant \frac{m^2r_{s}^6}{48M^2}.
\end{eqnarray}

We describe the system with dimensionless quantity $mr_{s}$,~$mM$ and $m^2\eta$
in accordance with the symmetry of the equation (3) in the form
\begin{eqnarray}\label{BHg}
r\rightarrow k r,~~~~ m\rightarrow m/k,~~~~ M\rightarrow k M,~~~~\eta\rightarrow k^2\eta.
\end{eqnarray}

From (16), we obtain a lower bound on the dimensionless coupling parameter $m^2\eta$ as
\begin{eqnarray}\label{BHg}
m^2\eta> \frac{m^4r_{s}^6}{48M^2}.
\end{eqnarray}
Below this bound, the static scalar field cannot exist
outside the horizonless neutral compact reflecting stars.
Above this bound, in the following section, we will numerically obtain
the scalar hairy configurations supported by
horizonless neutral compact reflecting stars.

\section{Static scalar hairy configurations supported by horizonless neutral compact reflecting stars}

As the scalar field equation (3) is of the second order,
we need values of $\psi(r_{s})$, $\psi'(r_{s})$, $mr_{s}$, $mM$
and $m^2\eta$ to integrate the equation from the star surface
to the infinity. Scalar reflecting surface conditions yield $\psi(r_{s})=0$.
With the symmetry $\psi\rightarrow k \psi$, one can take $\psi'(r_{s})=1$.
Fixing values of $mr_{s}$ and $mM$, we search for the proper $m^2\eta$
that can support a scalar field asymptotically decaying at the infinity.

We show the numerical results
with $mr_{s}=3.5$, $mM=1.5$ and various $m^2\eta$ in Fig. 1.
In the left panel, for $m^2\eta=132$, the scalar field increases
to be more positive at the infinity.
When we choose $m^2\eta=134$ in the right panel, the scalar field
decreases to be more negative in the larger r region.
So $m^2\eta=132$ and $m^2\eta=134$ are not related to the
bound-state scalar field.

\begin{figure}[h]
\includegraphics[width=180pt]{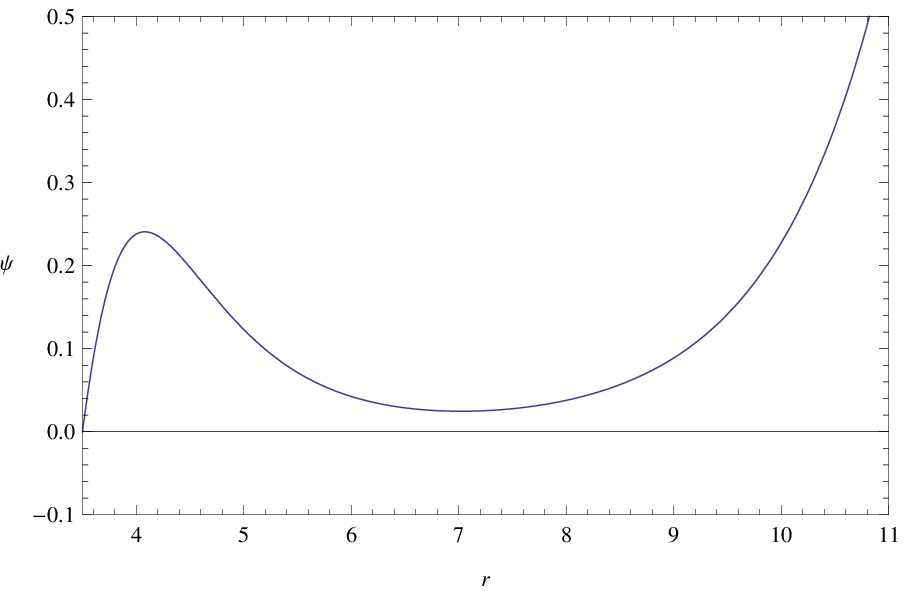}\
\includegraphics[width=180pt]{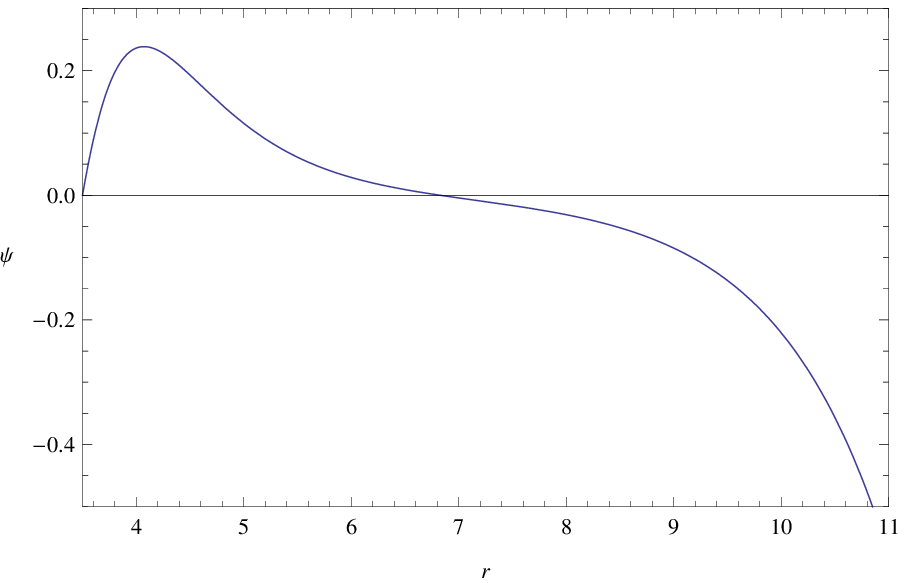}\
\caption{\label{EEntropySoliton} (Color online) We plot the scalar field $\psi(r)$
as a function of the coordinate $r$ with $mr_{s}=3.5$ and $mM=1.5$.
The left panel corresponds to $m^2\eta=132$ and the right panel is with
$m^2\eta=134$.}
\end{figure}

The mathematical solutions of equation (3) have the general asymptotic behavior
$\psi\thickapprox A\cdot\frac{1}{r}e^{-m r}+B\cdot\frac{1}{r}e^{m r}$ with $r\rightarrow \infty$.
We find $B>0$ in cases of $m^2\eta<133.002493$ and $B<0$ in cases of $m^2\eta>133.002493$,
also in accordance with results in Fig. 1.
It implies there is a critical value of $m^2\eta$ corresponding to $B=0$,
which leads to physical solutions with asymptotically decaying behaviors
$\psi\varpropto \frac{1}{r}e^{-m r}$ at the infinity.
With more detailed calculations, we obtain a discrete value $m^2\eta\thickapprox133.002493 $, which
corresponds to the scalar field satisfying $\psi(\infty)=0$.
We plot the scalar field with $mr_{s}=3.5$, $mM=1.5$
and $m^2\eta\thickapprox133.002493 $ in Fig. 2.
As shown in the picture, the scalar field approaches zero in the far region.
In the case of black holes, as explicitly shown in Refs. \cite{SGB2,SGB3}, in the
linearized regime, the non-minimally coupled
scalar hair is also characterized by a discrete resonant
set of the non-trivial coupling parameter.
It should be noted that the continuous spectrum found
numerically in \cite{SGB2,SGB3} belongs to non-linear
(self-gravitating) field configurations.

\begin{figure}[h]
\includegraphics[width=220pt]{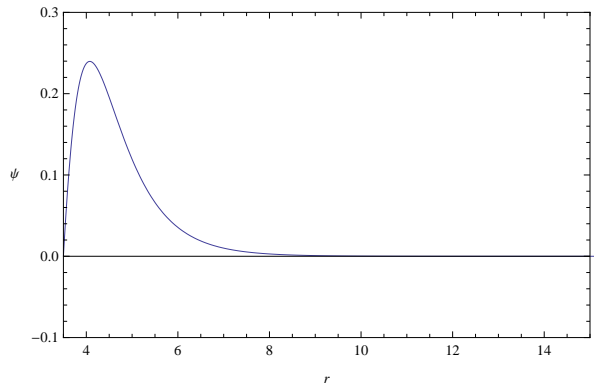}\
\caption{\label{EEntropySoliton} (Color online) We show behaviors of the scalar field function $\psi(r)$
with respect to the coordinate r for $mr_{s}=3.5$,~$mM=1.5$ and $m^2\eta=133.002493$.}
\end{figure}

In the following, we disclose effects of model parameters $mr$ and $mM$
on the dimensionless scalar-Gauss-Bonnet coupling parameter $m^2\eta$, which can support
the existence of bound-state scalar field hairs. In Table I, we show $m^2\eta$
with respect to different $mM$ in the case of
$mr_{s}=3.5$ and in Table II, we show values of $m^2\eta$ by choosing various $mr_{s}$ with $mM=1.5$.
We find that $m^2\eta$ decreases as a function of $mM$
and $m^2\eta$ becomes larger if we choose a larger value of $mr_{s}$.
In particular, according to the data in Table I, $m^2\eta$ is almost a linear function with respect to
$mM$.

\renewcommand\arraystretch{2.0}
\begin{table} [h]
\centering
\caption{The coupling parameter $m^2\eta$ with $mr_{s}=3.5$ and various $mM$}
\label{address}
\begin{tabular}{|>{}c|>{}c|>{}c|>{}c|>{}c|>{}c|}
\hline
$~mM~$ &~1.50~& ~1.55~& ~1.60~& ~1.65~& ~1.70\\
\hline
$~m^2\eta~$ & ~133.002493~ & ~122.364486~& ~112.393858~& ~102.719683~& ~92.479122\\
\hline
\end{tabular}
\end{table}

\renewcommand\arraystretch{2.0}
\begin{table} [h]
\centering
\caption{The coupling parameter $m^2\eta$ with $mM=1.5$ and various $mr_{s}$}
\label{address}
\begin{tabular}{|>{}c|>{}c|>{}c|>{}c|>{}c|>{}c|}
\hline
$~mr_{s}~$ & ~3.1~& ~3.3~& ~3.5~& ~3.7~& ~3.9\\
\hline
$~m^2\eta~$ & ~65.351225~& ~95.872104~& ~133.002493~& ~179.588365~& ~237.794024\\
\hline
\end{tabular}
\end{table}

\section{Conclusions}

We studied the formation of scalar field hairs in the background of
asymptotically flat horizonless neutral compact stars.
We considered a static scalar field coupled to the Gauss-Bonnet invariant.
At the star surface, we took the scalar reflecting condition.
With analytical methods, we obtained lower bounds
on the scalar-Gauss-Bonnet coupling parameter
as $m^2\eta> \frac{m^4r_{s}^6}{48M^2}$,
where $\eta$ is the coupling parameter, m is the scalar field mass,
M is the star mass and $r_{s}$ represents star radii.
Below the lower bound, static scalar fields cannot exist
or there is no hair theorem.
And above the bound, we numerically obtained discrete
coupling parameters $m^2\eta$, which can support static scalar hairs outside
horizonless neutral compact reflecting stars.
Moreover, we disclosed effects of model parameters on the discrete coupling parameter.

\begin{acknowledgments}

This work was supported by the Shandong Provincial Natural Science Foundation of China under Grant
No. ZR2018QA008. This work was also supported by a grant from Qufu Normal University of China under Grant
No. xkjjc201906.

\end{acknowledgments}

\end{document}